\documentclass{sig-alternate-2013}
\usepackage{color}
\usepackage{url}
\usepackage{verbatim} 
\usepackage{subfigure} 
\usepackage{multirow}
\usepackage{algorithm}
\usepackage[noend]{algorithmic}
\usepackage{xspace}
\usepackage{graphicx}
\usepackage{epsfig}
\usepackage{amsmath}
\usepackage{amssymb}
\usepackage{times}
\usepackage{chngcntr}
\usepackage{booktabs}

\permission{Copyright is held by the International World Wide Web Conference Committee (IW3C2). IW3C2 reserves the right to provide a hyperlink to the author's site if the Material is used in electronic media.}
\conferenceinfo{WWW'14,}{April 7--11, 2014, Seoul, Korea.}
\copyrightetc{ACM \the\acmcopyr}
\crdata{978-1-4503-2744-2/14/04}

\newcommand{\hide}[1]{}
\newcommand{\xhdr}[1]{\vspace{1.7mm}\noindent{{\bf #1.}}}

\newcommand{\denselist}{ \itemsep -2pt\topsep-10pt\partopsep-10pt }

\graphicspath{{./FIG/}}

\begin{document}

\title{Engaging with Massive Online Courses}

\newcommand{\sups}[1]{\raisebox{1ex}{#1}}
\numberofauthors{1}
\author{
\alignauthor Ashton Anderson \hspace{0.5cm}
Daniel Huttenlocher \hspace{0.5cm}
Jon Kleinberg \hspace{0.5cm}
Jure Leskovec\\
\affaddr{
\hspace{5mm}Stanford University\hspace{1.1cm}
Cornell University \hspace{1.2cm}
Cornell University \hspace{0.5cm}
Stanford University
} \\
\email{
ashton@cs.stanford.edu \hspace{0.5cm}
\{dph, kleinber\}@cs.cornell.edu \hspace{0.5cm}
jure@cs.stanford.edu}
}

\maketitle

\begin{abstract}

The Web has enabled one of the most visible recent developments in education---the deployment of massive open online courses. 
With their global reach and often staggering enrollments, MOOCs have the potential to become a major new mechanism for learning. Despite this early promise, however, MOOCs are still relatively unexplored and poorly understood. 

In a MOOC, each student's complete interaction with the course materials takes place on the Web, thus providing a record of learner activity of unprecedented scale and resolution. In this work, we use such trace data to develop a conceptual framework for understanding how users currently engage with MOOCs. We develop a taxonomy of individual behavior, examine the different behavioral patterns of high- and low-achieving students, and investigate how forum participation relates to other parts of the course.

We also report on a large-scale deployment of badges as incentives for engagement in a MOOC, including randomized experiments in which the presentation of badges was varied across sub-populations. We find that making badges more salient produced increases in forum engagement.

\end{abstract}

\vspace{3mm}

\noindent
{\bf Categories and Subject Descriptors:}
H.2.8 [{\bf Database management}]: Database applications---{\em Data mining}.


\section{Introduction}
\label{sec:intro}

Massive open online courses, or MOOCs, have recently garnered
widespread public attention for their potential as a new
educational vehicle.
There are now multiple MOOC platforms (including the edX consortium, Coursera and Udacity) offering hundreds of courses, some of which have had hundreds of thousands of students enrolled.
Yet, despite their rapid development and the high degree of interest they've received, we still understand remarkably little about how students engage in these courses. 
To reason about MOOCs we
generally apply our intuitions from university-level courses in the
offline world, thinking of students who enroll in a course for credit
or as an auditor, and who participate on an ongoing basis over a 10-14
week time period. 
Yet there is relatively little quantitative evidence
to support or refute whether such intuitions hold in the case of MOOCs.

Understanding how students interact with MOOCs is a crucial issue because it affects how we
evaluate their efficacy and how we design future online courses.
For students who treat MOOCs like
traditional courses,
which run at a fixed pace over a fixed time period, it makes
sense to talk about students ``falling behind'' or ``dropping out.''
But for students who might treat MOOCs as online reference works or textbooks,
a completely different set of expectations would apply, in which the
natural interaction style may consist of bursts of asynchronous engagement and selective sampling of content.
For students who might use MOOCs to
sharpen and test their skills in an area, 
it would be reasonable
to see them undertaking portions of the 
course work without ever viewing lecture content.

Despite the high level of interest in both
the popular press and recent academic literature,
there have been relatively few quantitative studies of
MOOC activity at a large scale, and relatively little understanding of
different ways in which students may be engaging with MOOCs (in Section \ref{sec:related} we review recent
work in the research literature.) 
Without this understanding
it has been hard to rigorously evaluate either
optimistic claims made about student experience in MOOCs,
or concerns raised about low rates of completion.
And to the extent that different types of student behaviors have been anecdotally identified in MOOCs, it has been hard to assess
how prevalent they are.

In this paper we propose a framework for understanding how students
engage with massive online courses, based on 
quantitative investigations of
student behavior in several large Stanford University courses offered on
Coursera (one of the major MOOC platforms).  Students in these courses
can engage in a range of activities---watching lectures, taking quizzes
to test their understanding, working on assignments or
exams, and engaging in
a forum where they can seek help, offer advice, and have
discussions. 

We find first of all that each student's activities in a course
can be usefully described by one of 
a small number of {\em engagement styles},
which we formalize in a taxonomy
based on the relative frequency of certain activities 
undertaken by the student.
These styles of engagement are quite different from one another,
and underscore the point that students display a small number of
recurring but distinct patterns in how they engage with an online course.
We also consider the issue of performance and grades in the course,
and show that thinking about grades in terms of different styles of
engagement can shed light on how performance is evaluated.

Our second main focus in this paper is the course forums, and
how to increase engagement in them.
They provide interesting patterns of interaction,
in which students engage not just with the course material 
but with each other.
To shift the ways in which students engage in the forums, we designed two large-scale interventions.
In particular, we report on our deployment of badges in 
the discussion forum of one of the largest MOOCs, 
with the badges
serving as incentives to increase forum activity.
We find through a randomized experiment that different ways of
presenting the badges---to emphasize social signals and progress toward 
milestones---can have an effect on the level of engagement.

We now give an overview of these two themes---the styles of engagement,
and the interventions to modify engagement.

\subsection*{\bf A taxonomy of engagement styles}

In characterizing the predominant styles of engagement, we consider the
two fundamental activities of (i) viewing a lecture and 
(ii) handing in an assignment for credit.
(Additional activities including ungraded quizzes and forum participation
will be considered later in the paper, but won't be explicitly used
in categorizing engagement.)
One basic attribute of any given student's behavior is the extent
to which her overall activity is balanced between these two modalities.

\begin{figure}[t]
    \centering
    \includegraphics{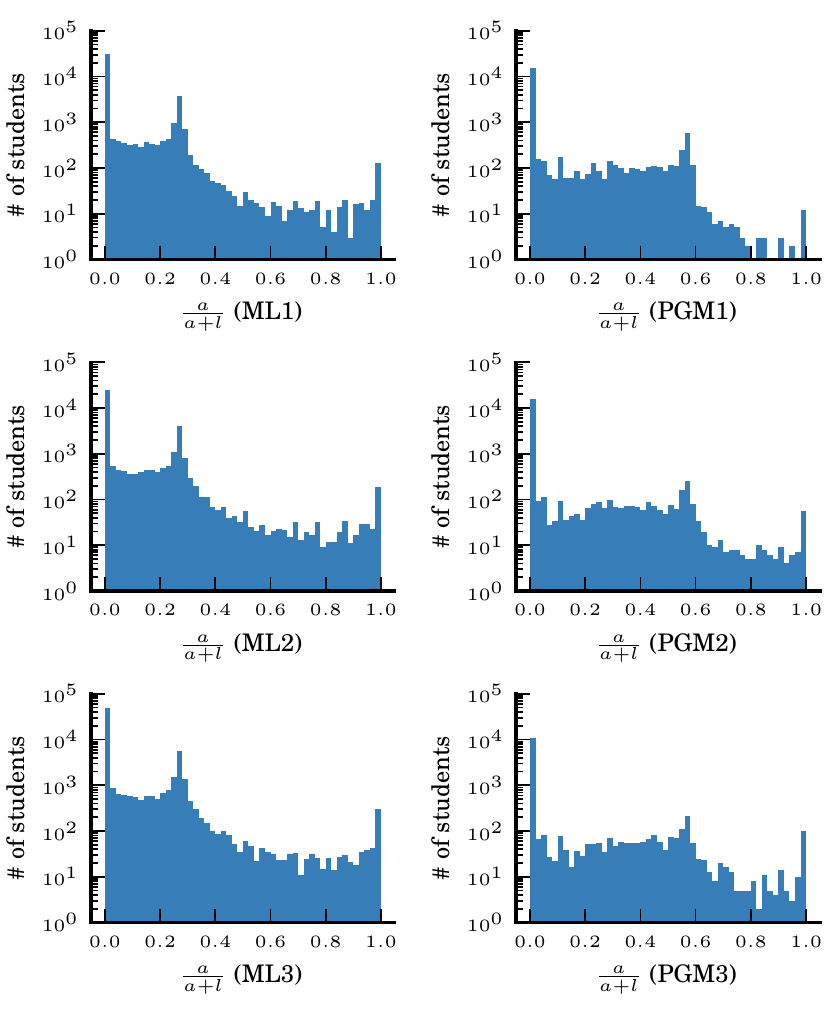}
    \vspace*{-0.35in}
    \caption{Distribution of {\em assignment fraction} (ratio of assignments, $a$, to overall activity of assignments and lectures, $a+l$). \vspace*{-0.35in}
    }
    \label{fig:assign_lecture}
\end{figure}

A natural way to address this question of balance between activities
is to compute a student's
{\em assignment fraction}: of the total number of lectures and assignments
completed by the student, what fraction were assignments?
Thus, a student with an assignment fraction of $0$ only viewed lectures,
while a student with an assignment fraction of $1$ only handed in 
assignments, without viewing any of the course content.
We computed the assignment fraction for each student in six Coursera
classes: three successive offerings of Machine Learning (which 
we name ML1, ML2, and ML3) and three successive offerings of
Probabilistic Graphical Models (which we name PGM1, PGM2, and PGM3).
Figure \ref{fig:assign_lecture} shows histograms of the number
of students with each assignment fraction in these six courses.
It is striking that the histogram for each course has three natural
peaks: a left mode near the fraction 0, a right mode near the fraction 1, 
and a central mode in between. 
This suggests three natural styles of engagement, one clustered
around each mode.

\begin{enumerate}
\vspace{-1mm}
\denselist
\item {\em Viewers}, in the left mode of the plot, 
primarily watch lectures, handing in few if any assignments.
\item {\em Solvers}, in the right mode, primarily hand 
in assignments for a grade, viewing few if any lectures.
\item {\em All-rounders}, in the middle mode, 
balance the watching of lectures with the handing in of assignments.
\vspace{-1mm}
\end{enumerate}

The number of distinct engagement styles is larger, however, due to
two other types of students who are not apparent from the three
modes of the histogram.
First, a subset of the students in the left mode are in fact downloading lectures
rather than viewing them on the site; this distinction is important because 
someone who only downloads content may or may not ever actually
look at it.  We therefore separately define a fourth style of engagement:
\begin{enumerate}
\vspace{-1mm}
\denselist
\item[4.] {\em Collectors}, also in the left mode of the plot, 
primarily download lectures, handing in few assignments, if any. 
Unlike Viewers they may or may not be actually watching the lectures.
\vspace{-1mm}
\end{enumerate}

Finally, there are students who do not appear in the plot, because they 
undertook very few activities:
\begin{enumerate}
\vspace{-1mm}
\denselist
\item[5.] {\em Bystanders} registered for the course but their total activity
is below a very low threshold.
\vspace{-1mm}
\end{enumerate}

These form our five styles of engagement: Viewers, Solvers, All-rounders, 
Collectors, and Bystanders.
While there are not sharp boundaries between them, we see
from the simple plot of relative activity levels that there are
natural behavioral modes that correspond to these cases. 
Moreover, we can formalize precise versions of these styles simply
by defining natural thresholds on the $x$-axis of the histograms in
Figure \ref{fig:assign_lecture} (to define clusters around the three modes), 
then separating out downloading
from watching lectures (to resolve the Collectors from the Viewers) 
and separating out students with very low total activity levels
(to identify the Bystanders).

Already from this simple taxonomy we can see that intuitions about
styles of student engagement from the offline world do not map
precisely onto the world of MOOCs. For instance, those just
``viewing'' lectures are not necessarily the same as auditors, because
if they are downloading rather than streaming they are quite possibly 
collecting the lectures for future use rather than watching them, much
as one would download an online textbook or other resource for
possible future use. 
And we find several other
fine-grained distinctions even within the different types,
although we do not separately distinguish them as distinct styles
of engagement.
For example, a subset of the All-rounders 
first hand in all the assignments and then download all the lectures. 
Such students are in a sense first behaving as Solvers, then 
subsequently as Collectors, but are certainly not participating in any way
that one might recognize as common in an offline course.

Moreover, this range of engagement styles shows that while the issue of 
students ``dropping out'' of MOOCs points to a genuine and important 
distinction in types of student activity, it is arguably a distinction being
made at too superficial a level.
Indeed, even asking whether a student ``completes'' an online course
is a question already based on the assumption 
that there is a single notion of completion.
While our taxonomy is based on activity traces and does not record
a student's internal motivations or intentions, we see that students
are reaching very different kinds of completion in their interactions
with these courses.
In particular, 
the relatively large numbers of Viewers shows that 
focusing primarily on those who engage with both the lecture content and the
graded assignments, as the majority of students do in offline courses,
overlooks a relatively large portion of those who are actively engaged
in MOOCs. Solvers similarly represent a style
of engagement that is different from the underlying intuition in this area;
many of them are likely
students who have previously learned the material or are
learning it elsewhere.  Both these types of students are presumably
getting value from the course, based on the fact that many complete
most of the assignments or view (or download) most of the lectures.

\xhdr{Engagement and Grades}
Our analysis of engagement styles also relates to the issue of students'
performance and grades; here too we find that one must
be careful in adapting intuitions from off-line courses.
First, we find in all the classes that most students receive a grade
of zero; however, the fact that many Viewers spend a non-trivial 
amount of time watching lectures means that a grade of zero should not be
equated with a failure to invest any effort in the course.
Second, we see an intriguing difference between the ML and PGM classes;
the PGM classes, which are more challenging, include students who do
all the coursework but have wide variation in their grades (as one sees
in challenging off-line courses), whereas in the ML classes we
find that a student's grade has a more linear relationship to the number
of assignments and lectures.  This suggests a useful distinction between
MOOCs in which a student's grade is a reflection primarily of effort expended
(as in ML), versus differential mastery of the work handed in (as in PGM).

\vspace{-1mm}
\subsection*{\bf Forums and Badges}

The second main focus in our work here 
is on the forums in online courses, and the design of mechanisms to increase activity in them.

Just as much of the discussion about student behavior on MOOCs
has been based on intuitions imported from other domains such as off-line
teaching, much of the reasoning about online course forums to date
has proceeded by analogy with forum-like modalities 
in other online settings.
The challenge in drawing such analogies, however, is that forums are
used in a very diverse set of contexts across the Web.
Should we think of an online course forum as behaving like
a discussion forum, where people engage in back-and-forth interaction;
or like a question-answering site, where some people come with questions
and others try to answer them; or like a product review forum,
where many people each individually react to a specific topic?

Our analysis shows that threads in the course forums proceed in 
a relatively ``straight-line'' fashion; each thread grows much more through
the arrival of brand-new contributors than through repeated
interaction among the initial contributors.  
Moreover, we find an interesting pattern in which the activity
level and average grade of the student making the initial post is
substantially lower than that of 
the students making subsequent posts.
This suggests that we may be observing a dynamic in which 
better students are helping out others in the class by
joining the threads they initiate.

\xhdr{Deploying Badges in the Forums}
We now discuss our work on designed interventions
to shift the level of student activity.
We did this in the context of forum activity, 
introducing badges into the forum for the third 
offering of Coursera's Machine Learning class (ML3).
Our badges were based on milestones for reaching certain activity levels
based on contributing to threads,
reading content, and voting on content.
When a student reached one of these milestones she received a badge.

It is interesting to note first of all that 
the ML3 course had significantly higher engagement compared
to earlier offerings of Machine Learning (ML1 and ML2) 
in which badges were not used.
While we cannot necessarily establish that this is due to the
introduction of badges, it is the case that the first two offerings
of the Machine Learning course were extremely similar to each other in
their forum behavior, and the third was quite distinct:
the distribution of forum contributions in ML3 developed a much
heavier tail, and a group of high-volume forum contributors emerged
who followed up much more actively on initial posts.
Moreover, the changes in ML3 were much more prominent on forum activities
where badges were introduced, and much less pronounced 
where they weren't. 

We were also interested in whether different ways of making the badges
salient to the students could produce different incentives and hence
different levels of forum activity, so we developed
a randomized experiment that presented the badges
differently to different sub-populations of the students.
As we discuss in Section \ref{sec:badgeexp}, making the
badges more salient produced an aggregate increase in forum activity;
the strongest effect came from a design that made a student's
own progress toward the badge visible and explicit,
but we also saw effects from more social mechanisms for creating
salience---displaying a student's current set of badges next
to her or her name for others to see.
It is striking that these relatively subtle differences produced
a non-trivial effect on forum contribution; students had knowledge of
and access to the same badges in all cases, and the only difference
was the extent to which the badge was emphasized in the user interface.

\xhdr{Data}
As noted above, our data comes from 
six Stanford classes offered on Coursera: three successive offerings
of Machine Learning (ML1-3) and three of Probabilistic Graphical Models
(PGM1-3).
ML3 is the course in which we
added badge-based incentives to the forum.
Table~\ref{tab:data} summarizes the number of actions of each type in each of the three offerings of the two courses.

\begin{table}[t]
\centering
\small
\tabcolsep=0.15cm
\begin{tabular}{lrrrrrrr}
\toprule
        Class & Students & HWs & Quizzes & Lectures & Posts & Start \\
        \midrule

        ML1 & 64,536 & 432,052 & 1,486,566 & 3,222,074 & 15,274 & 4/2012 \\
        ML2 & 60,092 & 488,554 & 1,563,301 & 3,066,189 & 15,763  & 8/2012 \\
        ML3 & 112,897 & 681,569 & 2,076,354 & 4,742,864 & 32,200 & 4/2013 \\
        PGM1 & 30,385 & 398,314 & 794,290 & 1,564,87 & 14,572 & 3/2012 \\
        PGM2 & 34,693 & 210,199 & 427,209 & 1,059,464 & 7,044 & 9/2012 \\
        PGM3 & 25,930 & 172,539 & 337,657 & 686,899 & 4,320 & 7/2013 \\
    \bottomrule
    \end{tabular}
    \caption{Basic course statistics. Posts are forum posts and Start is when the class started.}
    \label{tab:data}
\end{table}

\section{Patterns of Student Activity}
\label{sec:students}

\begin{table}[t]
\centering
\small
\tabcolsep=0.11cm

\begin{tabular}{lrrrrr}
\toprule
        Class & Bystander & Viewer & Collector & All-rounder & Solver \\ \midrule
ML1 & 28,623 (.47) & 15246 (.25) & 8,850 (.15) & 8,067 (.13)  & 378 (.01) \\
ML2 & 27,948 (.49) & 13,920 (.21) & 7,314 (.11) & 9,298 (.19)   & 550 (.01) \\
ML3 & 62,020 (.54) & 24,411 (.21) & 15,282 (.13) & 13,417 (.12) & 786 (.01) \\
PGM1 & 13,486 (.47) & 6,742 (.23) & 6,147 (.21) & 2,365 (.08) & 25 (.00) \\
PGM2 & 22,767 (.62) & 6,689 (.18) & 5,727 (.16) & 1,507 (.04) &  116 (.00) \\
PGM3 & 15,920 (.61) & 4,816 (.19) & 3,772 (.15) & 1,287 (.05) & 157 (.01) \\
 \bottomrule
    \end{tabular}

    \caption{Number (fraction) of students of different types.}
    \label{tab:types}
    \vspace{-3mm}
\end{table}

\xhdr{Engagement styles}
We begin by putting numerical estimates on the five styles of
engagement outlined in the introduction.
We will base our estimates on the structure of
Figure \ref{fig:assign_lecture},
since the three modes of the depicted distributions were the initial motivation behind our five categories: the left mode is comprised of a
mixture of Viewers and Collectors; the middle mode of All-rounders;
and the right mode of Solvers. Finally, students who only took very few actions are Bystanders.

Clearly there are not perfectly sharp distinctions between these
categories, but we can produce a working approximation by simply
dividing up the $x$-axis in 
Figure \ref{fig:assign_lecture}
into three intervals around the three modes.
More concretely, we first 
define numerical thresholds $c_0 \geq 1$ and $0 < \theta_0 < \theta_1 < 1$.
Then, for a given student who did $a$ assignment questions and consumed $l$ lectures, 
we call them:
\begin{itemize} 
    \denselist
    \item a \emph{Bystander} if $(a + l) \leq c_0$; otherwise, they are
    \item a \emph{Viewer} or \emph{Collector} if $a / (a + l) \leq \theta_0$;
    depending on whether they primarily viewed or downloaded lectures, respectively, 
    \item an \emph{All-rounder} if $\theta_0 < a / (a + l) < \theta_1$;
    \item a \emph{Solver} if $a / (a + l) \geq \theta_1$.
\end{itemize}
To produce the five sets, we choose the threshold $c_0 = 2$, 
$\theta_0$ to be the midpoint between the left and center mode,
and $\theta_1$ to be the midpoint between the center and right mode.
The resulting sizes of the five categories are shown in 
Table \ref{tab:types}.  (The partition appears to be
relatively robust numerically, in that fairly different approaches
to dividing the students into these five categories produce 
roughly comparable numbers.)
As noted earlier, although the Solvers
are a small category in relative terms, it is nevertheless remarkable
that several hundred such students should be present in these courses.

\begin{figure}[t]
    \centering
    \vspace{-2mm}
    \includegraphics{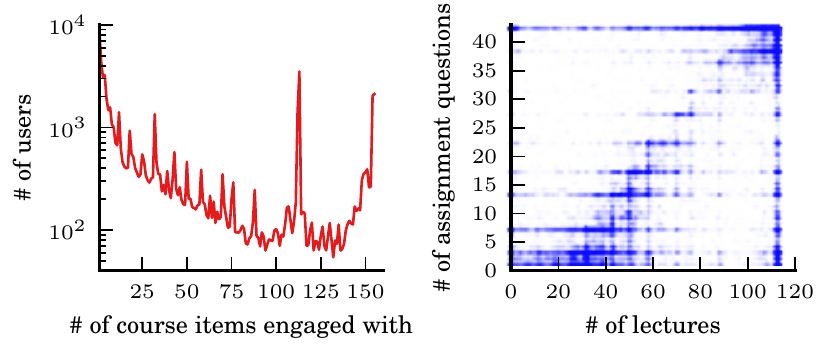}
    \caption{(a) Distribution over number of course items (lectures and assignment questions) engaged with in ML3; (b) Scatter plot of the same data.}
    \label{fig:numactions}
\end{figure}

These distinct modes of behavior show up in other ways of 
looking at student activity.
Figure~\ref{fig:numactions}a shows simply the distribution of 
the total number of items (assignment questions and lectures) engaged with in the course.
One immediately notices two large spikes in the curve, located
at the total number of lectures and the total number of lectures
plus assignment questions, respectively.

We can also look at variations in these patterns of activity.
Figure~\ref{fig:numactions}b
depicts the following scatter plot for ML3 (plots for the other courses are similar): for each student who has attempted $a$ assignment questions and consumed $l$ lectures, we put a point at $(a,l)$.  
Different styles of engagement occupy different parts of the
$(a,l)$ plane, but several other things stand out across all the courses
as well.  First, there is a ``dense part'' of the scatter plot corresponding
to a diagonal line of positive slope: in aggregate, $a$ and $l$ 
grow monotonically in one another.
Second, the horizontal stripes correspond to a collection of students
who all stopped after a particular assignment question.
And third, there is a corresponding lack of vertical stripes;
this suggests that students do not tend to stop at clearly
defined break-points in the sequence of lectures, as they do 
in the sequence of assignments.

\begin{figure}[t]
    \centering
    \vspace{-1mm}
    \includegraphics{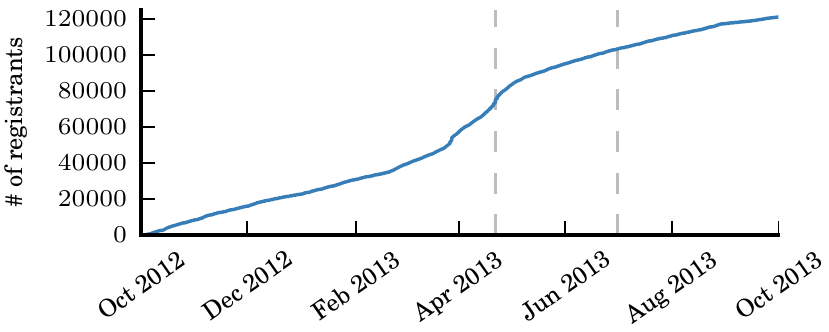}
    \caption{CDF of registration times in ML3. Vertical bars indicate the course's start and end dates.}
    \label{fig:engage_cdf}
\end{figure}

\xhdr{Time of interaction}
Our classification of students into five engagement styles is based on the total number of assignment questions they attempted and lectures they consumed. However, we also find that when a student interacts with a course is an important correlate of their behavior.
In the Coursera classes we study, students could register months in advance, much like many traditional courses. But unlike most offline courses, registration was often left open until months after the Coursera classes ended. Figure~\ref{fig:engage_cdf} shows that a significant fraction of students made use of this policy: only about 60\% of the students registered before the class officially began, and 18\% registered after it ended. This implies that a significant portion of the students are interacting with the class solely after it ends. We call these students \emph{archaeologists}. Precisely, a student is an archaeologist if her first action in the class is after the end date of the class (note that a student can register for the class at any time and still be an archaeologist, the only criterion is for her first action---and thus all her actions---to come after the end date). We keep this classification of students orthogonal from the engagement styles described above; what actions students take, and when they take them, are separate types of behavior. 
That so many students are interacting with courses after they end is an unexpected way in which MOOCs differ from traditional classes. 

\begin{figure}[t]
    \centering
    \includegraphics{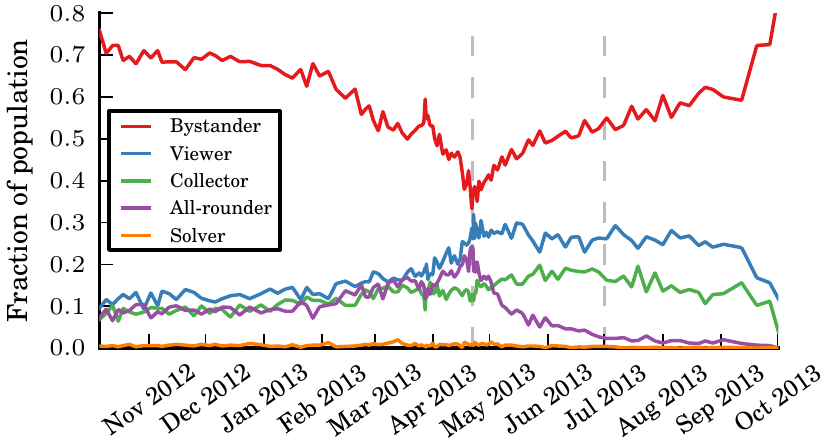}
    \caption{How the distribution of engagement styles varies with registration time. Vertical bars indicate the course's start and end dates.}
    \vspace{-2mm}
    \label{fig:engage-reg}
\end{figure}

A student who registers months in advance may have very different motivations and intentions from one who registers the day before it begins, or one who signs up months after it ends. In Figure~\ref{fig:engage-reg}, we plot the distribution of engagement styles as a function of registration time. The effect of registration time on engagement is strikingly large: for example, the fraction of Bystanders is as high as 70\% for students who sign up six months early, then sinks to just 35\% around the class's start date, and rises again to 60\% in the months after the class ends. We also see that late registrants are more likely to be Viewers and Collectors than early registrants are.

One can also examine the distribution of engagement styles as a function of first action time. Doing this, we discover that there is a crucial moment in determining students' engagement style: a significant amount of distributional mass abruptly shifts from All-rounders to Bystanders and Collectors the day after the first assignment is due.

\section{Grades and Student Engagement}
\label{sec:grade}

Having considered the basic styles and temporal patterns of engagement, we turn to an investigation of the grades that students receive for their work. In particular, we are interested in quantifying the relation between a student's grade and the way she engages with the course. 

We find that student engagement patterns are qualitatively similar among different versions of the same class, but there are interesting differences between ML and PGM. Thus, without loss of generality, in this section we will focus on the second iterations of both classes, ML2 and PGM2. 

\begin{figure}[t]
    \centering
    \includegraphics{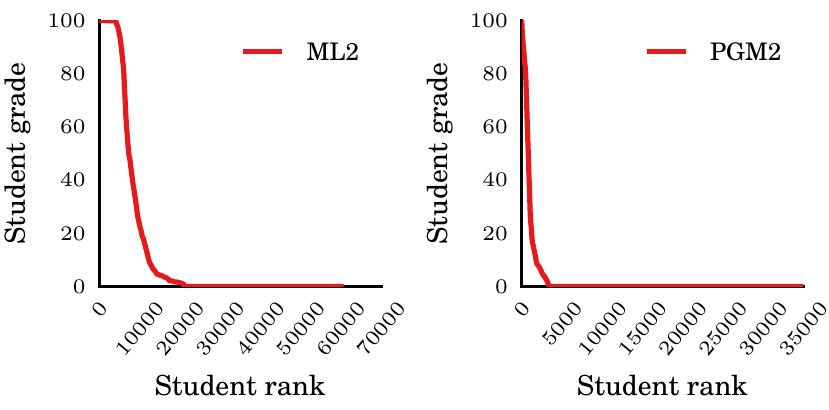}
    \caption{Final course grade distribution in ML2 and PGM2.}
    \label{fig:grade}
\end{figure}

\xhdr{Grade distribution}
First we examine the overall final grade distribution in the two classes, shown in Figure~\ref{fig:grade}. Notice that the grade distributions in the courses are heavily skewed; out of the 60,000 students who registered for ML2, two-thirds of them ($\approx$40,000) get a final grade of 0, 10\% of students ($\approx$5,000) achieve a perfect score, and the remaining 20\% of students ($\approx$10,000) receive a grade in between these two extremes. 
This breakdown is stable across all the versions of ML.
The distribution of grades in PGM2 is even more skewed: there is no large contingent of students with a perfect grade (only two students achieve this distinction), and only 10\% (3,300 out of 35,000 students) achieve a non-zero grade.

\begin{figure}[t]
    \centering
    \includegraphics{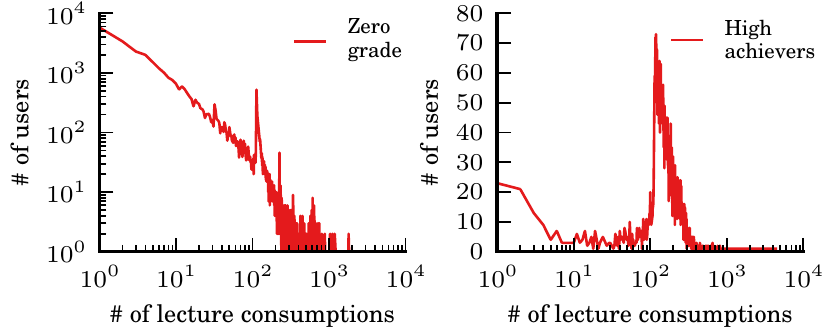} 
    \caption{Lecture watching behavior of students with final course grade of 0 (left) and 100 (right). The plots shows data for ML3 but we observe very similar behavior in all courses.}
    \label{fig:lectureviewing}
\end{figure}

In both classes there is a large number of students who achieve a score of zero in the course. However, this doesn't mean that these students are not engaging or putting effort in the course. In fact, many of the zero-grade students are Viewers who spend non-trivial amounts of time watching lectures. Figure~\ref{fig:lectureviewing}a shows the distribution of lecture views for zero-grade students. About 50\% of the registered students watch at least one lecture, and 35\% watch at least 10. The spike in the figure at $x=$ 120 corresponds to the total number of lectures in the class.

\begin{figure}[t]
    \centering
    \includegraphics{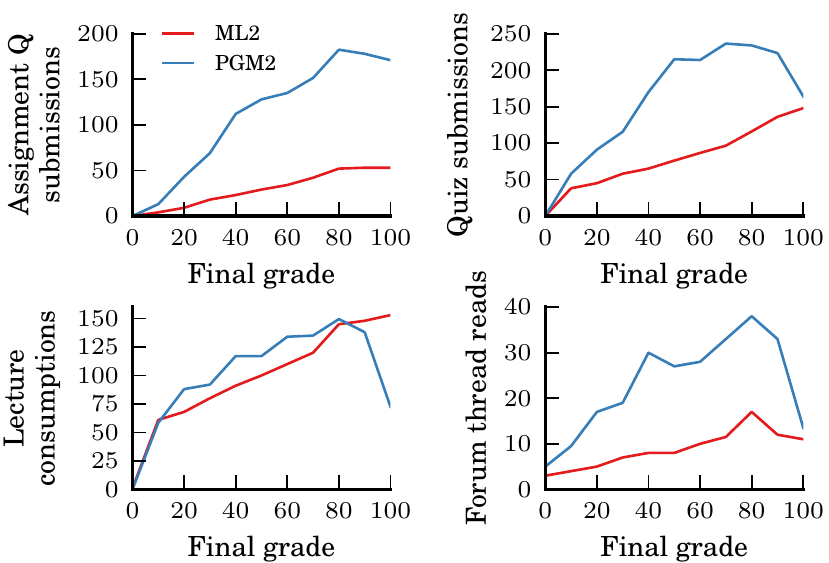}
    \caption{Median number of actions of students with a given final grade in PGM2 and ML2.}
    \label{fig:actionsgrade}
\end{figure}

\xhdr{Student's final grade and her engagement with the course}
We now investigate how a student's grade is related to her engagement and activity levels.
Here we think of the final grade as an independent variable; our goal is not to predict a student's grade from her activity but rather to gain insight into how high-grade and low-grade students distribute their activities differently across the site.

Figure~\ref{fig:actionsgrade} plots the median number of actions 
of a given type users take as a function of their final grade. 
Overall, the grade is generally proportional with their activity. In ML2, the median number of actions of a given type (assignment submissions, quiz submissions, lectures viewed, forum thread views) linearly increase with the student's final grade. 

PGM2 deviates from this general trend in an interesting way: the linear relationships only hold up to certain point. Class activity increases until a grade of around 80\%, but then decreases. For example, PGM students with near-perfect grades watch around the same number of lectures as students who got 20\%. Perhaps this is due to PGM being a highly technical course where students who have seen the content before have an easier time, while others seem to struggle. Finally, both courses deviate from the linear trend in forum reading. We observe that students with perfect grades read less on the forums than those with lower grades (Figure~\ref{fig:actionsgrade}d).

\xhdr{Behavior of high-achievers}
We now examine the activity of ``high-achievers'', the students who scored in the top $10^{\rm th}$ percentile of the class.

A key trait of high-achievers is that they consume many lectures. Most of them have more lecture watches than there are lectures in the course (indicating some re-watching behavior); although the mode of the distribution is at exactly the number of videos in ML2, the distribution is skewed to the right (see Figure~\ref{fig:lectureviewing}b).
In the plot we also clearly observe the population of Solvers, who watch very few (or no) lectures.

\begin{figure}[t]
    \centering
    \includegraphics{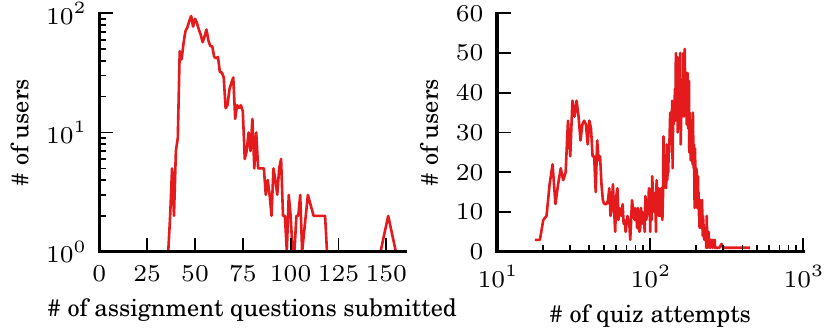}
    \caption{Number of handed-in assignments (left) and quizzes (right) for high-achievers.}
    \label{fig:perfectstudents}
\end{figure}

While lecture watching is characteristic of high-achievers, the number of assignment question attempts, however, is surprisingly variable (Figure~\ref{fig:perfectstudents}a). 
Out of a total of 42 assignment questions, the mean number of submissions high-achievers handed in was 57 (mode 42), while some submitted more than 200. 
Although it is no surprise that in order to finish the course with a perfect grade one has to submit all the assignments, it is interesting to observe the bimodal distribution of the number of quizzes students submitted (Figure~\ref{fig:perfectstudents}b). Here we observe that most students submitted around 120 quizzes, while the first mode of the distribution is at around 30. This is consistent with the explanation that the population of near-perfect students is composed of two subgroups: Solvers, who perhaps already know the material, and All-rounders, who diligently watch the lectures, finish the quizzes, and do assignments.

\section{Course Forum Activity}
\label{sec:forum}

\begin{figure}[t]
    \centering
    \includegraphics{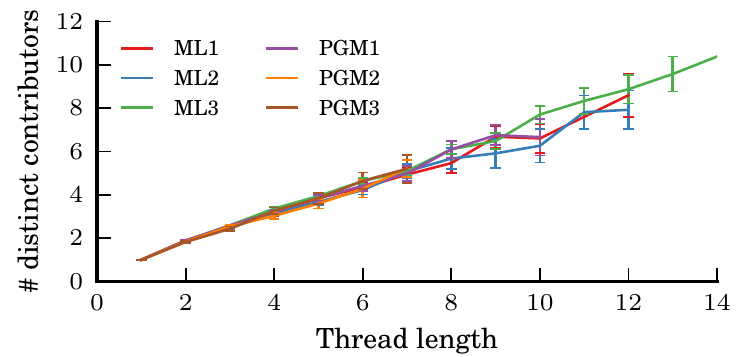} 
    \caption{Number of distinct contributors as a function of thread length.}
    \label{fig:forum-distinct03}
\end{figure}

We now move on to our second main focus of the paper, the forums, which provide a mechanism for students to interact with each other. 
Because Coursera's forums are cleanly separated from the course materials, students can choose to consume the course content independently of the other students, or they can also communicate with their peers.

Following our classification of students into engagement styles, 
our first question is a simple one: which types of students visit the forums?
To answer this, we compute the distribution of engagement styles for the population of students who read at least one thread on ML3 (shown in the top row of Table~\ref{tab:forum-engage}). 
The representation of engagement styles on the forum is significantly different from the class
as a whole, with more active students over-represented on the forum;
for example, 
Bystanders comprise over 50\% of registered students but only
10\% of the forum population.

We also compute the fraction of each engagement style present
on the forum (the bottom row of Table~\ref{tab:forum-engage}). It
is striking that 90\% of All-rounders are forum readers,
meaning that the two populations heavily overlap. While numerically the forum is used by a small fraction of the full population of registered
students, this is a superficial measure; using our engagement taxonomy
it is apparent that a large
majority of the most engaged students are on the forum.

\begin{table}[t]
\tabcolsep=0.12cm
\centering
    \begin{tabular}{cccccc}
    & Bystander & Viewer & Collector & All-rounder & Solver \\
    \midrule
    $P(S|F)$ & 0.106 & 0.277 & 0.192 & 0.408 & 0.017 \\
    $P(F|S)$ & 0.050 & 0.334 & 0.369 & 0.894 & 0.648 \\
    \end{tabular}
\caption{How engagement styles are distributed on the ML3 forum. $P(S|F)$ is probability of engagement style given forum presence (reading or writing to at least one thread); $P(F|S)$ is probability of forum presence given engagement style.}
\label{tab:forum-engage}
\end{table}

\xhdr{The composition of threads}
The forum is organized in a sequence of {\em threads}: 
each thread starts with an initial post from a student,
which is then potentially followed by a sequence of further posts.
Threads cover a variety of topics: discussion of course content,
a question followed by proposed answers, 
and organizational issues 
including attempts by students to find study groups they can join.

Forum threads are a feature of a wide range of Web sites---social networking sites, news sites, question-answer sites,
product-review sites, task-oriented sites---and they are used quite differently across domains.  
Thus one has to be careful in adapting existing intuitions
about forums to the setting of online courses---in principle
it would be plausible to conjecture that the forum might be a place
where students engage in back-and-forth discussions about course
content, or a place where students ask questions that other students answer,
or a place where students weigh in one after another on a class-related issue.
Our goal here is to develop an analysis framework
that can clarify how the forums are in fact being used.

In particular, we'd like to address the following questions:
\begin{itemize}
\denselist
\item Does the forum have a more
conversational structure, in which a single student may 
contribute many times to
the same thread as the conversation evolves, or a
more straight-line structure, in which most students contribute
just once and don't return?
\item Does the forum consist of high-activity students who initiate
threads and low-activity students who follow up, or are the threads
initiated by less central contributors and then picked up by
more active students?
\item How do stronger and weaker students interact on the forum?
\item Can we identify features in the content of the posts that
indicate which students are likely to continue in the course
and which are likely to leave?
\end{itemize}

The course forums contain many threads of non-trivial
length, and we ask whether these threads are long because
a small set of people are each contributing many times to a long
conversation, or whether they are long because a large number of
students are each contributing roughly once.

As a first way to address this question, 
we study the mean number of distinct contributors in a thread of length $k$,
as a function of $k$.
If this number is close to $k$, it means that many students are contributing;
if it is a constant or a slowly growing function of $k$, then 
a smaller set of students are contributing repeatedly to the thread.

We find that
the number of distinct contributors grows linearly in $k$
(see Figure \ref{fig:forum-distinct03}):
a thread with $k$ posts has roughly $2k/3$ distinct contributors.
Moreover, this slope is markedly consistent across all six courses
in our data.
The linear growth in distinct contributors
forms an interesting contrast with discussion-oriented sites;
for example, on Twitter and Facebook, the number of distinct
contributors in a thread of length $k$ grows sublinearly in $k$
\cite{kumar-conversations-kdd09,backstrom-wsdm13}.

Now, it is possible for the actual number of distinct contributors
to exhibit two modes; for example, long threads on Facebook have
this multi-modal behavior, as long conversational threads among
a few users co-exist with ``guest-book'' style threads that
bring in many users \cite{backstrom-wsdm13}.
In our domain, however, we find 
a single mode near the mean number of distinct users;
long conversational threads with very few contributors are extremely 
rare in these course forums.

\xhdr{Properties of thread contributors}
Even if we know the forum is dominated by threads with many
distinct contributors, there are still several possible dynamics
that could be at work---for example, a top-down mechanism in which a high-activity forum user
starts the discussion, or an initiator-response mechanism in which
a less active user begins the thread and more active users
continue it.

\begin{figure}[t]
    \centering
    \includegraphics{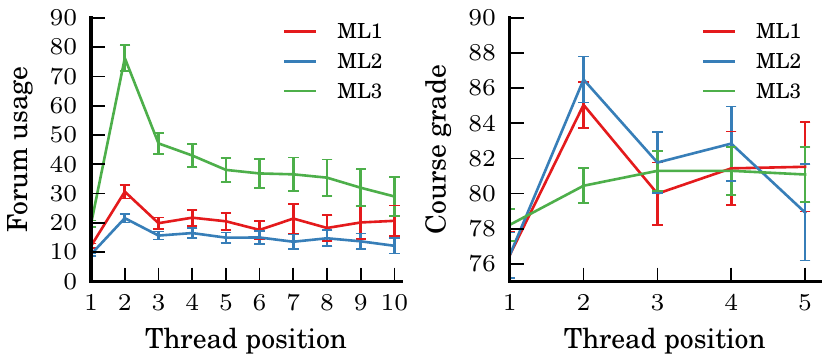}
    \caption{(a) Average forum usage as a function of thread position, (b) Average grade as a function of thread position (ML).}
    \label{fig:forum-activity01}
    \label{fig:forum-grade-pos-ml}
\end{figure}

One way to look at this question is to plot a student's 
average forum activity level as a function of her position in 
the thread---that is, let $f(j)$ be the lifetime number of forum 
contributions of the student in position $j$ in the thread, 
averaged over all threads. 
We do this in Figure \ref{fig:forum-activity01}a, finding a
clear initiator-response structure in which the second contributor
to the thread tends to have the highest activity level, 
reacting to the initial contribution of a lower-activity student.
This effect is especially pronounced in ML3,
where badges were present; it suggests a hypothesis that 
as students sought badges, 
a common strategy was to react quickly to new threads
as they emerged and make the next contribution.

In the ML classes, we also see a reflection of this distinction
between the thread initiator and later contributors when we
look at the grades of the thread contributors.  
In particular, let $g(j)$ be the final course grade of the student
in position $j$ in the thread, averaged over all threads.
We find (Figure \ref{fig:forum-grade-pos-ml}b) that the
initiator of a thread tends to have a significantly lower
grade than the later contributors to the thread.
In the first and second ML classes, this was particularly true
of the second contributor; the effect was reduced in the third
ML class, where the second contributor tended to be extremely
active on the forum but less distinctive in terms of grades.
Overall, however, this suggests a valuable structure to the forum,
with stronger students 
responding to the posts of weaker students, as one would hope to find
in a peer-interaction educational setting.

In the PGM classes, the same pattern holds for forum activity
(the function $f(j)$), but the grade function $g(j)$ is essentially flat:
the thread initiator has roughly the same grade as the later contributors.
It would be interesting to relate this to the other grade distinctions
between the ML classes and PGM classes that we saw in Section \ref{sec:grade}.

\xhdr{Lexical analysis}
Finally, we consider the content of the posts themselves.
As discussed above, a basic question is to understand whether
we can analyze student engagement based on what they write in
their posts---can we estimate a student's eventual activity level from their early forum posts?

One way to make this question precise is as follows.
Consider all forum posts for the first two weeks of the course;
and for each word $w$ that occurred a sufficient number of times, 
let $h(w)$ be the final number of assignments
handed in by a student who uses the word $w$ in their post,
averaged over all posts in the first two weeks that contain $w$.
Do certain words have particularly high or low values of $h(w)$
relative to the overall distribution?

In all six classes, there was a consistent pattern
to the extremes of $h(w)$.
For illustration, 
Table~\ref{tab:forum-words01} shows the words achieving the
highest and lowest $h(w)$, over all words $w$ occurring
at least 150 times.
The high end of $h(w)$ contains 
a number of terms suggesting the
author is already familiar with some course terminology, and 
hence potentially some of the course content.

Perhaps more interestingly, 
the low end of $h(w)$ was consistently characterized by two points.
First, many of the words with low $h(w)$
are associated with students using the forum to form study groups or
to find study partners.
Second, many of the words with low $h(w)$ are non-English words.
This finding suggests that the platform
could potentially be improved by 
more effective mechanisms for helping students form
study groups, and making the course more accessible to speakers
of languages other than English.

\begin{table}
\centering
\small
\begin{tabular}{p{0.75in}|p{2.1in}}
\toprule
Highest number of assignments & values, your, error, different, x, using, matrix, cost, 
function, gradient \\ 
\midrule
Lowest number of assignments & que, de, computer, study, am, interested, I'm, me, hello, new\\ 
\bottomrule
\end{tabular}
\caption{For each word $w$, we look at the average final number of assignments
handed in by students who use 
the word $w$ in a post in the first two weeks.
The table depicts the words whose usage is associated with the highest
and lowest average number of assignments submitted.
}
\label{tab:forum-words01}
\end{table}

\section{A Large-Scale Badge Experiment}
\label{sec:badgeexp}
We've discussed many facets of the behavior we observe in our data:
the various types of engagement styles students exhibited, the grades
they received, and their activity on the forums. Now we report on two
large-scale interventions we carried out on the third run of Coursera's
Machine Learning class (ML3). First, we designed and
implemented a badge system on the forums; and second, we ran a
randomized experiment that slightly varied the presentation of the
badges to different groups of users. As a result of these two
interventions, we find compelling evidence that introducing a badge
system significantly increased forum participation, and that even
small changes in presentation can produce non-trivial differences in
incentives---thereby helping to shed light on how badges generate
their incentive effects. We'll first discuss the badge system we
implemented, and then conclude with the randomized experiment.

\subsection{Badge system}

In previous work, we studied the role of badges as incentives both through
a theoretical model and an analysis of user behavior on Stack Overflow,
a question-answering site with a thread structure similar to what
we have observed in the course forums here \cite{anderson-badges}.
We studied ``milestone'' badges, where users win badges once they perform a pre-specified amount of some activity.
We found that as users approached the milestone for a badge,
they ``steered'' their effort in directions that 
helped achieve the badge. In other words, the badge was
producing an incentive that guided their behavior. 

Badges are a similarly prominent feature of many thriving online forums, 
and are generally viewed as producing incentives for participation.
The extent to which they actually do so, however, remains unclear. 
Are badges indeed significant drivers of engagement, 
or are they mainly incidental to behavior on the site?
As Coursera was interested in increasing forum engagement, we had the
opportunity to design and implement a badge system for ML3's
forums, which provided an ideal setting to study this
question.

\xhdr{Badge types}
Our badge system operated entirely on the ML3 forums; there was no
connection to course-related actions like doing assignments or
watching lectures. We primarily used milestone badges, 
the same type we studied in our previous work, but we also introduced
some other types described below.
One of the main design principles suggested by our prior work is that
a suite of several, less-valuable badges targeting the same action,
carefully placed in relation to each other, can be more effective as a group
than one single ``super-badge''. Following this principle, as
well as some previous industry implementations, we designed four badge
levels---bronze, silver, gold, and diamond---which were associated with 
increasing milestones, and were correspondingly difficult to attain.

We decided to award badges for some actions but not others. We awarded
simple \emph{cumulative} badges for reading certain numbers of
threads, and voting on certain numbers of posts and comments, but we
awarded no such badges for authoring posts or threads. This was to
avoid incentivizing the creation of low-quality content solely to win
badges. To discourage ``low-quality votes'', users had to wait a small period of time
between votes, to prevent users from quickly voting on a succession of posts or comments at random.

Instead of cumulative badges for authoring posts and threads, we instead implemented \emph{great achievement} badges to incentivize high-quality content, which were awarded for authoring posts or threads that were up-voted a certain number of times by other forum members. The bronze, silver, gold, and diamond levels corresponded to increasing numbers of up-votes as milestones.

We also created \emph{cumulative great achievement} badges to reward users who consistently authored high-quality content, as judged by their peers. We called a post or thread ``good'' if it received at least three up-votes, and bronze, silver, good, and diamond badges were awarded for authoring certain numbers of good posts and threads.

Finally, we created a number of one-time badges for different
purposes. ``Community Member'' and ``Forum Newbie'' simply welcomed
users to the main course and course forums respectively (and
introduced users to the badge system). The ``Early Bird'' badge was
given for participation in the forums in the early days of
the class, to help the forums get off to a good start, and the
``All-Star'' badge was given to users who were active on the forums
during every week of the class. All the badges are shown in
Figure~\ref{fig:badges}c.

\xhdr{Effects of the badge system on forum engagement}
As a first contrast between the forums with and without badges, we analyze
the distribution of the number of actions users took on ML3, and compare
it to the same distributions in previous runs of the class, ML1 and
ML2, neither of which had a badge system in place. If badges
incentivized users to be more engaged in the forums, then we should
observe a shift in the distribution of the number of actions in ML3
toward a heavier tail---indicating that more users took more
actions relative to the number of users who took fewer actions.
As we discuss more fully below, 
we cannot necessarily conclude that the differences
exhibited by ML3 relative to ML1 and ML2 are the result of the badge system,
but a close look at the distinctions between the courses provides evidence for the impact of badges.
(And in the next subsection we discuss the results of our
randomized experiment within ML3, where we could vary the conditions
more precisely.)

\begin{figure}[!tbp]
    \centering
    \includegraphics{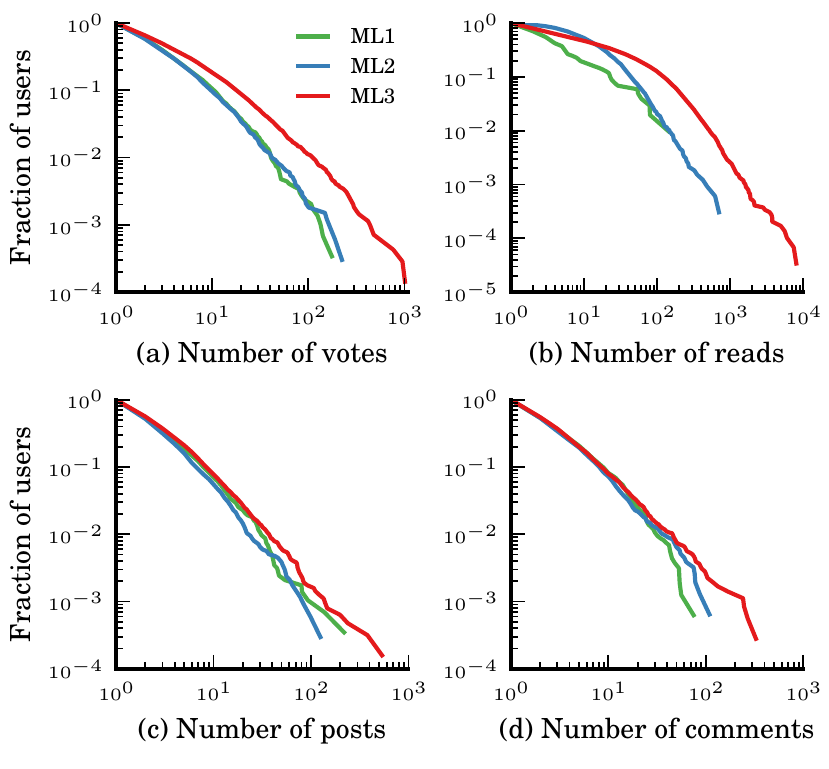}
    \caption{Normalized CCDFs of action counts. Engagement on actions with cumulative badges (voting, reading) is significantly higher in ML3 than in ML1 and ML2; engagement on actions without badges (posts, comments) is essentially the same across the three runs.}
    \label{fig:action_ccdfs}
\end{figure}


In Figure~\ref{fig:action_ccdfs}a, we show the complementary
cumulative distribution function (CCDF) on votes, where the point
$(x,y)$ means $y$ fraction of users voted at least $x$ times (as
a fraction of the total number of users who voted at least once in that
run of the class). We show the normalized values (the fraction of
students) instead of the absolute magnitudes (the number of students)
because the different runs of Machine Learning had different numbers
of students. By normalizing each curve by the total number of students
who voted at least once in that run, we enable comparisons between the
shapes of the distributions, which is what we are interested in.

First, observe that the distribution in ML3 is clearly more
heavy-tailed than the distributions of both ML1 and ML2, indicating
that a larger fraction of users voted many times in ML3 than in ML1
and ML2. This indicates that users were more engaged in
voting in ML3. 
For example, relative to the number of students who voted at least once,
the fraction of students who voted at least 100 times was 10 times larger
in ML3 than in the other two runs.
Second, notice that the distributions in ML1 and ML2
are essentially identical, despite these two runs of the course 
having been offered at different times.
This suggests that certain features of the distribution were stable
prior to the striking difference exhibited by ML3, in which badges were
offered.

Similarly, the distribution of threads viewed is strikingly different
in ML3 compared to ML1 and ML2 (see Figure~\ref{fig:action_ccdfs}b).
Again, relative to the number of students who read at least one thread,
the number of students who read at least 100 threads was 10 times larger
in ML3 than in the other two runs. Students were
substantially more engaged in viewing threads and voting on posts and
comments in the ML3 run than they were in ML1 and ML2.

What about the other forum actions that were available to students?
In Figure~\ref{fig:action_ccdfs}c, we show the distributions of the number of posts users authored. Although they vary slightly, the distributions are largely similar. The differences between ML3 and ML2 are on the same scale as the differences between ML1 and ML2. And as can be seen in Figure~\ref{fig:action_ccdfs}d, the differences in the distributions of the number of comments are even smaller. Thus, the actions that didn't have cumulative badges, authoring posts and comments, didn't show qualitatively significant differences in engagement between the three runs of the class. This is in stark contrast with the actions with cumulative badges, voting and reading threads, where users were much more engaged in ML3 than in ML1 and ML2.

Finally, we examine whether content quality was different in the version
of the course with badges,
by comparing the normalized CCDFs of votes per item.
The distributions are again very similar across classes---if
anything, the distribution in ML3 is slightly heavier than the other
two, suggesting posts were more likely to receive more votes in ML3.  
Thus, by this measure of voting, post quality didn't suffer in ML3.

The observational comparisons in this section aren't
sufficient to make a definitive claim that the badge system
we designed and implemented was responsible for the striking increase
in user engagement on the forum. There could be other
factors, unobservable by us, that are responsible. But there are several
consistent points in support of the hypothesis that our badge system
played an important role. First, we compared our results to two
separate controls, the engagement levels on two previous runs of the
class. These controls were always very similar to each other, which
both validates them as reasonable controls and renders less plausible
alternate hypotheses positing that different class runs are incomparable.
Second, and even more striking, is that the forum engagement on ML3 increased in
specific ways that completely paralleled the portions of
the course targeted by the badge system. 
The actions with cumulative badges, voting and thread reading, saw big
increases in engagement, whereas the actions without cumulative
badges, post and comment authoring, didn't significantly change.
Any alternate hypothesis thus requires some difference between ML3 and the
other two runs that didn't exist between ML1 and ML2, and furthermore this difference must hold for the targeted actions (voting and thread reading) but not for the control actions (post and comment authoring). 
We don't have natural alternate hypotheses for how these particular parts of
the course changed significantly while the others were unaffected. 

\subsection{Badge Presentation Experiment}
A big open question from our previous work was to understand how and why
badges were producing their incentive effects.
Were users viewing the badges as goals to be achieved for intrinsic
personal reasons?  Or were they viewed as signals of social status,
with the incentive effect correspondingly coming from the badge's
visibility to others?
These questions are important for determining how best
to design badge systems, but without the ability to experiment, 
there is no clear way to distinguish among these hypotheses.

We performed an experiment in which we 
randomly partitioned the student population of ML3
and presented them with badges in slightly different ways.
Some of the badge presentations were designed to amplify the 
personal goal-setting aspect of badges: the sequence of milestones
and the student's partial progress were highlighted and made salient.
Other aspects of the badge presentations were designed to amplify their
social function: students' badges were displayed next to their name,
creating a sense of social status that might have motivated them and/or other
students.

In the end, all of the modified presentations were quite subtle,
and none of them shifted the actual milestones for achieving the badges.
Thus, there is no a priori reason to be confident that they should
have had any effect at all.
But in fact, we will see that they produced non-trivial effects,
and for one of presentations in particular they produced quite 
a significant effect.

\xhdr{Badge treatment conditions}
We seek to study the effect of badges on student behavior and thus to investigate several dimensions in the badge design space. 
We aim to answer questions including:
\begin{itemize}
	\denselist
	\item What is the effect of the badges on forum participation? 
	\item Are students motivated more when they see their own badges, or does seeing others' badges have a larger impact?
	\item Does it help to keep the student informed about their future prospects for earning badges?
\end{itemize}

With the above questions in mind we designed the following three different treatment conditions, which are  outlined in Figure~\ref{fig:badges}:

\begin{figure}[t]
	\begin{tabular}{l}
	\hline
	{\bf (a) Top-byline: badge byline shown in header of every page:}\\
    \includegraphics[width=0.48\textwidth]{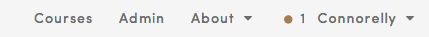} \\ 
    \\ \hline
	{\bf (b) Thread-byline: badge byline displayed on every forum post:}\\
    \includegraphics[width=0.48\textwidth]{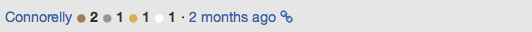} \\
	\\ \hline
    {\bf (c) Badge-ladder: badge ladder on a student's profile page:}\\
    \includegraphics[width=0.48\textwidth]{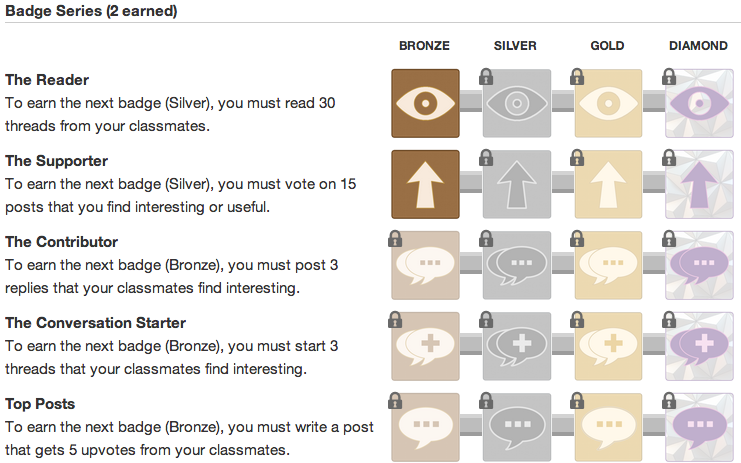} \\ 
    \\ \hline
    {\bf (d) Badge-ladder control: badge list on a student's profile page:}\\
    \includegraphics[width=0.25\textwidth]{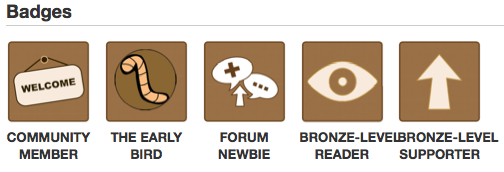} \\
    \end{tabular}
    \vspace{-4mm}
    \caption{Three badge experimental conditions (a,b,c) and the badge-ladder control (d).}
   \vspace{-4mm}
    \label{fig:badges}
\end{figure}

    {\em (a) Top byline}: In the header of a Coursera webpage a student saw a byline showing the counts of how many different levels of badges (bronze, silver, etc.) she has already won. Figure~\ref{fig:badges}a shows an example of the top byline of a student who earned 1 bronze badge. This treatment was aimed to test whether students change their behavior when the badges they already have is more visible to them. Students in the control group saw the same header without the badge byline.

    {\em (b) Thread byline} condition is similar to Top byline, except here every post on the forum is annotated with the byline of badges earned by the author of that post. In this experimental condition, badges are more visible, and are more strongly linked with user identity on the forum since names are always accompanied by badge counts (Figure~\ref{fig:badges}b shows an example). Students in the control group didn't see any badge bylines accompanying author names.

    {\em (c) Badge ladder} tested the effect of keeping
students informed about their progress towards badges and the badges they can win in the future. Figure~\ref{fig:badges}c shows an example of a badge ladder, where we see badges earned by the student, all other ``locked'' badges, as well as conditions for obtaining them. In this experimental condition, upon winning a badge the student would receive a pop-up (a ``toast'') with a congratulatory message and, crucially, information about how many more actions were needed to win the badge at the next level (e.g. gold, after winning a silver badge). In the control group, the student could only see the badges she obtained so far on her profile page (see Figure~\ref{fig:badges}d), and the congratulatory message displayed upon a badge win wasn't accompanied with 
information about future badges.

Overall, we created $2^3=8$ experimental buckets of students: one for each possible combination of control and treatments for the three independent experiments. 

\xhdr{Effect of badge treatment conditions}
Now we turn to measuring the treatment effects and their corresponding statistical significance. As our primary goal was to increase engagement and student participation in the forums, we take the total number of actions on the forum to be our main outcome of interest (where actions are posts and views).

With this outcome we need to take care when measuring the treatment effects. The number of forum actions is distributed according to a heavy-tailed distribution (in all of the classes), which implies that a small number of users are responsible for a large fraction of the total actions taken on the forums. Thus comparing the mean numbers of actions of users in the treatment and control groups is subject to these extreme outliers, and the median numbers of actions will be too low, thus limiting our statistical power. Since we are mainly interested in whether users in the treatment groups were generally more active on the forums than the users in the control groups, we adopt the Mann-Whitney U test, a rank-sum non-parametric test that is relatively robust to the outliers we expect given our underlying heavy-tailed distribution.

\begin{table}[t]
\tabcolsep=0.2cm
\centering
    \begin{tabular}{ccc}
    Top byline & Thread byline & Badge ladder \\
    \midrule
    0.095 & 0.095 & 0.036 \\
    \end{tabular}
\caption{Mann-Whitney p-values for the three treatments.}
\label{tab:pvalues}
\end{table}

The significance values by the end of the class were 0.095 for top-byline, 0.095 for thread-byline, and 0.036 for badge ladder: 
thus, badge ladder is significant at the 5\% level, 
and top-byline and thread-byline are significant at the 10\% level. 
Between the three treatments we tested, badge-ladder clearly had the most significant effect.
Top-byline and thread-byline were less significant but still performed better than we'd expect from null treatments, 
suggesting that these treatments also revealed some signal in engagement. 

There are many potential reasons 
why the badge-ladder was successful, 
but it is natural to suppose
that making the next badge milestone clear was important in keeping users 
engaged on the forums.  Assembling the badges in coherent groups made it easy to give users targets to shoot for. Similarly, the byline treatments could have been effective for various reasons. The thread byline treatment made badges more visible, and perhaps attached badges to one's identity on the site, potentially creating a social effect. 
Further disentangling what make badges successful will make it easier to deploy 
them successfully in other domains in the future.

We find it striking that even the subtleties in the design 
of the badging system, let alone the presence of the badges themselves, 
have significant effects on forum engagement and activity. 
These results thus provide evidence that badges can be 
sufficiently powerful motivators that even relatively small changes in their presentation can change the resulting effects they have on users. The variety of the treatments we tried, and the fact that they were all significant to varying degrees, is suggestive of the richness of underlying psychological mechanisms that could be behind badges' effectiveness. Following up on the hints offered by our large-scale randomized experiment and further elucidating these mechanisms is an exciting direction for future work.

\section{Related Work}
\label{sec:related}

As noted earlier, there have to date been very few quantitative studies of
student behavior on MOOCs.  Two recent papers in this theme have been
a study using edX \cite{Breslow} and an analysis
of sub-populations of students in MOOCs using a Coursera course
\cite{Kizilcec}.
The first of these papers focuses on the amount of time that students 
spend engaged in various activities, 
as well on demographic information about the students,
with a different focus than our work here.
The second considers patterns of actions taken by the
students, but unlike the engagement styles we define, their
patterns are based on the timing of a student's actions
relative to the course schedule.

On the topic of online forums for education, there has been a significant
amount of work in settings predating the current growth of MOOCs.
Davies and Graff studied forums in online learning,
finding that participation in an online discussion forum was linked
to higher academic performance \cite{davies-graff}.
The benefits of discussion forums as a supplement to learning
has been explored in a number of other settings as well \cite{deslauriers,palmer,vonderwell}.

Finally, our use of badges in the forum experiment follows up 
on recent work that analyzes the effects of badges on user behavior
\cite{anderson-badges,easley-badges}.
In contrast to these recent papers, our work here does not analyze
trace data from an existing badge system, but instead designs a
new badge system and incorporates a randomized experiment designed
to test how badges produce their effects.

\section{Conclusion}
\label{sec:conclusion}

In this paper we have taken some initial steps toward characterizing
the ways in which students engage with massive open online courses (MOOCs),
and we have explored methods for increasing the level of student activity
in these settings.
Rather than thinking of MOOCs as online analogues of traditional
offline university classes, we found that online classes come with
their own set of diverse student behaviors. 
In an analysis based on a set of large MOOCs offered by Stanford University, 
we identified five different categories of student behavior,
all quite distinct from one another.

To shift patterns of student engagement in these courses, and in
the course discussion forums in particular,
we deployed a system of badges designed to produce incentives for activity
and contribution.
As part of this deployment, we performed
a large-scale randomized experiment in which we varied the ways in 
which badges were presented to different sub-populations of the students.
Even small variations in badge presentation had an effect on activity.

It seems clear that the proliferation of online courses in different formats
and at different scales will lead to a set of fundamental research questions
about the power of these new educational environments and the ways 
in which students engage with them.
We see the present work as one step in the exploration of these questions.
Among the many potential directions for future work, we mention the following:
predictive models of student behavior and grades; personalization and 
recommendation mechanisms to help increase user engagement and learning; 
identifying effective student behavior and developing methods that 
automatically recognize students who require help or are lacking
an understanding of a key concept; 
understanding and facilitating students' use of forums and discussion boards; 
and further exploring badges and other incentive mechanisms 
to not only increase student engagement but also to
help students learn material more effectively.

\xhdr{Acknowledgments} {\small We thank Andrew Ng, Daphne Koller, Pamela Fox, and Norian Caporale-Berkowitz at Coursera for their help with implementing the badge system and for sharing the data with us. Supported in part by a Google PhD Fellowship, a Simons Investigator Award, an Alfred P. Sloan Fellowship, a Google Research Grant, ARO MURI, DARPA SMISC, PayPal, Docomo, Boeing, Allyes, Volkswagen, Intel, and NSF grants IIS-0910664, CCF-0910940, IIS-1016099, IIS-1016909, CNS-1010921, IIS-1149837, and IIS-1159679.}
\vfill\eject
\nocite{*}
\bibliography{refs}
\bibliographystyle{abbrv}

\end{document}